\documentclass[draftcls,onecolumn]{IEEEtran}

\usepackage{amssymb,amsfonts}

\usepackage{graphicx}% Include figure files
\usepackage{dcolumn}% Align table columns on decimal point
\usepackage{bm}% bold math

\newcommand{\Z}{\mathbb{Z}}
\newcommand{\beq}{\begin{equation}}
\newcommand{\eeq}{\end{equation}}

\title{Relating Granger causality to directed information theory for networks of stochastic processes}

\author{ Pierre-Olivier Amblard$^{1,2}$ and Olivier J.J. Michel$^1$ \\
 $^1$ GIPSAlab/CNRS UMR 5216/ BP46, \\ 38402 Saint Martin d'H\`eres cedex, France \\
 $^2$ The University of Melbourne, Dept. of Math\&Stat. 
 Parkville, VIC, 3010, Australia\\
{\tt bidou.amblard@gipsa-lab.inpg.fr}  \\
{\tt olivier.michel@gipsa-lab.grenoble-inp.fr}
}

\begin{document}

\maketitle

%% PLACE YOUR ABSTRACT HERE
\begin{abstract}

This paper addresses the problem of inferring circulation of information between multiple stochastic processes. 
We discuss two possible frameworks in which the problem can be studied: directed information theory and Granger causality.
The main goal of the paper is to study the connection between these two frameworks. In the case of directed information theory, we stress the importance of Kramer's causal conditioning. This type of conditioning is necessary not only in the definition of the directed information but also for handling causal side information. We also show how directed information decomposes into the sum of two measures, the first one related to Schreiber's transfer entropy quantifies the dynamical aspects of causality, whereas the second one, termed instantaneous information exchange, quantifies the instantaneous aspect of causality. After having recalled the definition of Granger causality, we establish its connection with directed information theory. The connection is particularly studied in the Gaussian case, showing that Geweke's measures of Granger causality correspond to the transfer entropy and the instantaneous information exchange. This allows to propose an information theoretic formulation of Granger causality.

\end{abstract}

{\bf keywords}
directed information, transfer entropy, Granger causality, graphical models

%\newpage

\sloppy

\section{Introduction}

The importance of the network paradigm for the analysis of complex systems, in  fields ranging from biology and  sociology to  communication theory or computer science, gave rise recently to the emergence of new research interests referred to as network science or complex network \cite{BarrBV08,FranM07,Newm03}. Characterizing the interactions between the nodes of such a network is a major issue for understanding its global behavior and identifying its topology. It is customary assumed that nodes may be observed via the recording of (possibly multivariate) time series at each of them, modeled as   realizations of   stochastic processes (see  \cite{RaoHSE06,RaoHSE07} for examples in biology, or  \cite{LungS06,Spor07,BullS09,KrasSG04} for applications in neurosciences). The assessment of an interaction between two nodes  is then formulated as a interaction detection/estimation problem   between their  associated time series. Determining the existence of edges between given nodes (or vertices) of a graph may be reformulated in a graphical modeling inference framework \cite{Whit89,Laur96,Edwa00,Pear00,Laur01}. Describing connections in a graph requires to provide a definition for the interactions that will be carried by the edges connecting the nodes. Connectivity  receives different interpretations in the neuroscience literature for instance, depending on whether it is `functional', revealing some dependence, or `effective' in the sense that it accounts for   directivity \cite{JirsM07,Spor07}. This differentiation in the terms describing connectivity raises the  crucial issue of causality, that goes beyond the problem of simply detecting the existence or the strength of an edge linking two nodes. 

Detecting whether a connection between two nodes can be given a direction or two can be addressed by identifying possible `master-slave' relationships between nodes. Based on the measurements of two signals $x_t$ and $y_t$, the question is: `Does $x_t$ influences  $y_t$ more than $y_t$ influences $x_t$?'. Addressing this problem requires the introduction of tools that account for  asymmetries in the signals information exchanges. 

Granger and others investigated this question using the concept of causality \cite{Gran80,Gran88,Gewe82, Pear00} and emphasized that  interaction between two processes is relative to the set of observed nodes. Actually, the possible interactions of the studied pair of nodes with other nodes from the network may profoundly alter the estimated type of connectivity. This leads to fundamental limitations of pairwise approaches for multiply connected network studies. 
Many authors addressed the topic of inferring causal relationship between interacting stochastic systems under the restriction of linear/Gaussian assumptions. In \cite{Gewe82,Gewe84}  the author develops a general linear modeling approach in the time domain. A spectral domain definition of causal connectivity is proposed in \cite{KamiDTB01}, whose relationship with Granger causality is explored in \cite{Eich06}. However, all these techniques need to be extended or revisited to tackle nonlinearity and/or nonGaussianity.  

Information-theoretic tools provide a means to go beyond Gaussianity. Mutual information characterizes the information exchanged between stochastic processes \cite{CoveT93,Pins64}. It is however a symmetric measure and does not provide any insight on possible directionality. Many authors have managed to modify mutual information in order to obtain asymmetrical measures. These are for example  Saito and Harashima's transinformation  \cite{SaitH81,KamiHMS84,AlkhA08}, the coarse grained transinformation proposed by Palus {\it et al.} \cite{PaluKHS01,PaluV07}, Schreiber's transfer entropy  \cite{Schr00,KaisS02}. All these measures share common roots which are revealed using directed information theory. 

\subsection{Main contributions of the paper}

This paper is an attempt to make sense of and to systematize the various definitions and measures of causal dependence that have been proposed to date. Actually, we claim that  these measures can be reduced to directed information, with or without additional causal conditioning. 
Directed information introduced by Massey in 1990 \cite{Mass90} and based on the earlier results on Marko's bidirectional information theory \cite{Mark73}, is shown to be an adequate quantity to address the topic of causal conditioning within an information theoretic framework. Kramer, Tatikonda and others  have used directed information to study communication problems in systems with feedback \cite{Kram98,Tati00,VenkP07,TatiM09}. Although their work  aimed at  developing new bounds on the capacity of channels with feedback and optimizing directed information, most of their results allow better insight in causal connectivity problems for systems that may exhibit feedback.  

Massey's directed information will be extensively used to quantifying directed information flow between stochastic processes. We  show how directed information, which is intimately linked to feedback,  provides a nice answer to the question of characterizing directional influence between processes, in a fully general framework. A contribution of this paper is to describe the link between Granger causality and directed information theory, both in the bivariate and multivariate cases. It is shown that  causal conditioning plays a key role as its main  measure, directed information, can be used to assess causality, instantaneous coupling and feedback in graphs of stochastic processes. A main contribution is then a reformulation of Granger causality in terms of directed information theoretic concepts.

\subsection{Organization of the paper}
As outlined in the preceding sections, directed information plays a key role in defining information  flows in networks  \cite{Mark73,SaitH81,KamiHMS84,Mass90,Kram98,Tati00,TatiM09,RaoHSE06,RaoHSE07,Solo08,AmblM08,AmblM09}. 
Section \ref{dirinfo:sec} gives a formal development of directed information following earlier works of Massey, Kramer and Tatikonda \cite{Mass90,Kram98,Tati00}.
Feedback in the definition of directed information is revisited, together with its relation to Kramer's {\em causal conditioning} \cite{Kram98}. 
This paper extends these latter ideas and shows that  {\em causally conditioned directed information} is a means of measuring directed information in networks: 
it actually accounts for the existence of other nodes interacting with those studied.  The link between directed information and transfer entropy \cite{Schr00}  established in this section is a contribution of the paper.  
In section \ref{granger:sec} we present Granger causality  which  relies on  forward prediction. We particularly insist on the case of  multivariate time series.
Section \ref{DirInfoGranger:sec} is devoted to developing the connection between the present information theoretic framework and  Granger causality.  Although all results hold in a general framework explained in section \ref{defproposal:ssec}, a particular attention is given to  the Gaussian case. In this case  directed information theory and Granger causality are shown to lead to equivalent tools to assess directional dependencies (see also \cite{AmblM09}). This  extends similar recent results  independently obtained by Barnett {\it et. al.} \cite{BarnBS09} in the case of two interacting signals without  instantaneous interaction.  
An enlightening illustration of the interactions between three time series is presented in section \ref{threeprocesses:sec} 
for a particular Gaussian model.

\section{Measuring directional dependence}
\label{dirinfo:sec}

\subsection{Notations and basics}

Throughout the paper we consider discrete time, finite  variance $E[|x|^2] < +\infty$ stochastic processes. Time samples are indexed by $\Z$; $x_k^{n}$ stands for the vector  $(x_k, x_{k+1}, \ldots, x_n)$, whereas for $k=1$, the index will be omitted for the sake of readability. Thus we identify the time series $\{x(k), k=1,\ldots,n\}$ with the vector $x^n$. $E_x[.]$ will denote the expectation with respect to the probability measure describing $x$, whereas $E_p[.]$ will indicate that the expectation is taken with respect to the probability distribution $p$.

In all the paper, the random variables (vectors) considered are either purely discrete, or continuous with the added assumption that the probability measure is absolutely continuous with respect to the Lebesgue measure. Therefore, all derivations hereafter are valid for either cases. Note however that existence of limits will be in general assumed when necessary and not proved.

Let  $H(x^n)=-E_{x}[\log p(x^n )] $ be the entropy of a  random vector $x^n$ whose density is $p$. 
Let the conditional entropy be defined as $ 
H(x^n | y^n )=-E[\log p(x^n | y^n )] $.
The mutual information $I(x^n ; y^n)$  between vectors $x^n$ and $y^n$ is defined as \cite{CoveT93}: 
\begin{eqnarray}
I(x^n ; y^n) &=& H( y^n )-H(y^n| x^n) \nonumber \\ 
&=& D_{KL}\left( p(x^n ,y^n)\big|\big| p(x^n ) p(y^n)\right)
\end{eqnarray}
where $D_{KL}(p||q)= E_p[\log p(x)/q(x)] $ is the Kulback-Leibler divergence.  It is 0 if and only if $p=q$ almost everywhere and is positive otherwise.  The mutual information effectively measures independence since it is 0 if and only if  $x^n$ and $y^n$ are independent random vectors. As   $I(x^n ; y^n) = I( y^n ; x^n )$, mutual information  cannot handle directional dependence. 

Let $z^n$ be a third time series. It may be a multivariate process accounting for side information (all available observation but $x^n$ and $y^n$).  To account for  $z^n$, the conditional mutual information is introduced~: 
\begin{eqnarray}
 I(x^n ; y^n | z^n )  &= &E_z\big[ D_{KL}\big( p(x^n ,y^n| z^n)|| p(x^n | z^n ) p(y^m| z^n)\big)\big] \\
		&=& D_{KL}\big(p(x^n ,y^n , z^n) || p(x^n | z^n ) p(y^n| z^n) p(z^n) \big)
\end{eqnarray}
 $ I(x^n ; y^n | z^n ) $ is  zero if and only if  $x^n  $ and $y^n$ are independent {\em conditionally} to $z^n$. 
Stated differently,  conditional mutual information measures the divergence between the actual observations and those which would be observed under Markov  assumption $(x \rightarrow  z \rightarrow y)$. Arrows may be misleading here, as by reversibility of Markov chains, the equality above holds also for $(y \rightarrow  z \rightarrow x)$. This again emphasizes the inability of mutual information  to provide answers to the information flow directivity problem.

\subsection{Directed information}

 Directed information was introduced by Massey \cite{Mass90}, based on the previous concept of ``bidirectional information" of Marko \cite{Mark73}. Bidirectional information focuses  on the two nodes problem, but accounts for the respective roles of feedback and memory in the information flow.

\subsubsection{Feedback and memory}
Massey \cite{Mass90}   noted that the joint probability distribution   $p(x^n,y^n)$ can be written as a product  of two terms~: 
\begin{eqnarray}
\overleftarrow{p} (x^n|y^{n-1} ) &=  & \prod_{i=1}^n p(x_i | x^{i-1},y^{i-1} ) \nonumber \\
\overrightarrow{p}(y^n | x^n) &= &\prod_{i=1}^n  p(y_i | x^{i},y^{i-1}) \nonumber  \\
p(x^n,y^n )  & =&  \overleftarrow{p} (x^n|y^{n-1} ) \overrightarrow{p}(y^n | x^n) 
\label{factorisation:eq}
\end{eqnarray}
where for  $i=1$ the first terms are respectively $p(x_1) $ and $p(y_1|x_1)$. 
Assuming that $x$ is the input of a 
 system that creates $y$,  $\overleftarrow{p} (x^n|y^{n-1} )$ can be viewed as a characterization  of  feedback in the system. Therefore the name {\em feedback factor}: each of the factors controls the probability of the input $x$ at time $i$ conditionally to its past and to the past values of the output $y$.
Likewise, the term $\overrightarrow{p}(y^n | x^n)$ will be referred to as the {\em feedforward factor}. 
The factorization (\ref{factorisation:eq}) leads to some remarks:
\begin{itemize}
\item In the absence of feedback in the link from $x$ to $y$, one has
\beq
p(x_i \big| x^{i-1}, y^{i-1} ) = p(x_i \big| x^{i-1})  \mbox{ } \forall i\geq 2
\eeq 
or equivalently 
\beq
H(x_i \big| x^{i-1}, y^{i-1} ) = H(x_i \big| x^{i-1})  \mbox{ } \forall i\geq 2
\eeq
As a consequence~: 
\beq
 \overleftarrow{p} (x^n|y^{n-1} ) = p(x^n)
\eeq 
\item If  the feedforward factor does not depend on the past, the link is  memoryless~: 
\beq
 p(y_i \big| x_{i}) = p(y_i | x^{i},y^{i-1}) \mbox{ } \forall i\geq 1
\eeq
\item 
Let $D$ be the unit delay operator, such that $Dy_n=y_{n-1}$. We define $Dy^n= (0, y_{1}, y_{2},\ldots, y_{n-1}) $ for finite length sequences, in order to  deal with edge effects while maintaining constant dimension for the studied time series\footnote{The term $0$ in  $Dy^n= (0, y_{1}, y_{2},\dots, y_{n-1})$ indicates a wild card which plays no influence on conditioning, and makes sense as $y_0$ is not assumed observed.}. Then we have
\beq
\overrightarrow{p} (x^n|Dy^n )  = \overleftarrow{p} (x^n|y^{n-1} ) 
\label{backforward:eq}
\eeq
The feedback term in the link $x\rightarrow y$ is the feedforward term of the delayed sequence in the link $y\rightarrow x$. 

\end{itemize}

%%%%%%%%%%%%%%%%%%%%%%%%

%%%%%%%%%%%%%%%%%%%%%%%%
\subsubsection{Causal conditioning and directed information}
In \cite{Kram98}, Kramer introduced an original point of view, based upon the following remark. The conditional entropy is easily expanded (using Bayes rules) according to 
\beq
H\big(y^n \big| x^n\big)=   \sum_{i=1}^n H\big(y_{i} \big| y^{i-1} , x^n\big)
\eeq
where each term in the sum is the conditional entropy of $y$ at time $i$ given its past and {\em the whole observation of $x$}~:  
Causality (if any)  in the dynamics $x \rightarrow y$ is thus not taken into account. Assuming that  $x$ influences $y$ 
 through some unknown process, Kramer proposed that
the conditioning of $y$ at time $i$ should include $x$ from  initial time up to time $i$ only. He named this  {\em causal conditioning}, and defined {\em causal conditional} entropy as
\beq
H\big(y^n \big|\big|  x^n\big)=   \sum_{i=1}^n H\big(y_{i} \big| y^{i-1} , x^{i}\big)
\eeq
By plugging causal conditional entropy in the expression of mutual information in place of the conditional entropy, we obtain  a definition of {\em directed information}~:
\beq
I(x^n \rightarrow  y^n) = H\big(y^n ) - H\big(y^n \big|\big|  x^n\big)=   \sum_{i=1}^n I\big(x^{i} ; y_{i} \big| y^{i-1} \big)
\label{eq:directedInformation}
\eeq
 Alternately, Tatikonda's work  \cite{Tati00} leads to express directed information as a Kullback-Leibler divergence\footnote{The proofs rely on the use of the chain rule $I(X,Y;Z) = I(Y;Z|X) + I(X;Z)$ in the definition of the directed information.}~: 
 \begin{eqnarray}
I(x^n \rightarrow y^n ) &=& D_{KL} \left(  p(x^n,y^n) ||  \overleftarrow{p} (x^n|y^{n-1} ) p(y^n)\right)  \label{infodirKL:eq}\\
&=& E\left[  \log \frac{p( x^n|y^n)}{ \overleftarrow{p} (x^n|y^{n-1} )} \right]  \nonumber\\
&=&E\left[  \log \frac{ \overrightarrow{p}( y^n|x^n)}{ p(y^n )} \right] \label{infodirKLsnd:eq}  
\end{eqnarray}
The expression (\ref{infodirKLsnd:eq}) highlights the importance of the feedback term when comparing mutual information  with directed information:  $p(x^n)$ in the expression of  the mutual information is replaced by the feedback factor $\overleftarrow{p} (x^n|y^{n-1} )$ in the definition directed information. 

This  result allows the derivation of many (in)equalities rapidly. First, as a divergence, the directed information is always positive. 
Then, since
\begin{eqnarray*}
I(x^n \rightarrow y^n ) = E\left[ \log\Big(\frac{p(x^n,y^n)}{\overleftarrow{p} (x^n|y^{n-1} ) p(y^n) } \times \frac{p(x^n)}{p(x^n)}\Big)\right]
\end{eqnarray*}
Using equations    (\ref{backforward:eq}) and (\ref{infodirKLsnd:eq}) we get 
\begin{eqnarray}
-E\left[\log \frac{p(x^n)}{ \overleftarrow{p} (x^n|y^{n-1} )  }\right] = I( Dy^n \rightarrow x^n )
\end{eqnarray}
Substituting this result  into eq. (\ref{eq:directedInformation}) we obtain
\begin{eqnarray}
I(x^n \rightarrow y^n ) &=& I( x^n ; y^n ) + E\left[\log \frac{p(x^n)}{ \overleftarrow{p} (x^n|y^{n-1} )  }\right] \label{infomutdirdelay:eq}  \\
&=&  I( x^n ; y^n ) - \sum_i I(x_i ; y^{i-1} \big| x^{i-1} ) \nonumber \\
&=& I( x^n ; y^n )  -  I( Dy^n \rightarrow x^n ) \label{decompdi:eq}
\end{eqnarray}
Equation (\ref{decompdi:eq}) is fundamental as it shows how mutual information splits into the sum of a feedforward information flow $I(x^n \rightarrow y^n )$ and a feedback information flow $I( Dy^n \rightarrow x^n )$. 
In this absence of feedback,  $\overleftarrow{p}(x^n | y^n)  = p(x^n)$ and $I(x^n;y^n )=I(x^n \rightarrow y^n )$.  Equation  (\ref{infomutdirdelay:eq}) shows  that the mutual information is always greater than the directed information, since  $ I( Dy^n \rightarrow x^n )=\sum_i I(x_i ; y^{i-1} \big| x^{i-1} ) \geq 0 $. As a sum of positive terms, it is zero if and only if all the terms are zero~: 
\begin{eqnarray*}
I(x_i ; y^{i-1} \big| x^{i-1} ) = 0 \mbox{ } \forall i=2,\ldots,n
\label{infonofeedback:eq}
\end{eqnarray*}
or equivalently
\begin{eqnarray}
H(x_i \big| x^{i-1},y^{i-1}  ) = H(x_i \big| x^{i-1} ) \mbox{ } \forall i=2,\ldots,n
\label{entropienofeedback:eq}
\end{eqnarray}
This last equation  states  that
without  feedback,  the past of $y$ does not influence the present of $x$ when conditioned 
on its own past. 
Alternately, one sees that if eq. (\ref{infonofeedback:eq}) holds, then the sequence
$y^{i-1} \rightarrow x^{i-1}\rightarrow x_i $ forms a Markov chain, for all $i$: 
again, the conditional probability of  $x $  given its past does not depends on the past of $y$. 
Equalities (\ref{entropienofeedback:eq}) can be considered as a definition of the absence of feedback from $y$ to $x$.
 All this findings are summarized in the following theorem:

{\bf  Theorem:} (\cite{Mass90} and \cite{MassM05})
{\em 
 The directed information is less than or equal to the mutual information, 
 with equality   if and only if there is no feedback.} 
 %%%%%%%%%%%%%%%%%%%%%%%%%%%%%%

This theorem implies  that mutual information over-estimates  the directed information between two processes in the presence of feedback. This was thoroughly studied  in  \cite{Kram98,Tati00,VenkP07,TatiM09}, in a communication theoretic framework.

Summing the information flows in  opposite directions gives: 
\begin{eqnarray}
I(x^n \rightarrow y^n ) + I(y^n \rightarrow x^n ) &=& E\left[  \log \frac{ p(x^n,y^n)}{\overleftarrow{p} (x^n|y^{n-1} ) p(y^n)}  +\log \frac{ p(x^n,y^n)} {\overleftarrow{p} (y^n|x^n ) p(x^n) }\right] \nonumber \\
&=& I(x^n ;y^n ) + E\left[  \log \frac{ \overrightarrow{p} (y^n|x^n ) }{ \overleftarrow{p} (y^n|x^n ) } \right]  \nonumber \\
&=& I(x^n ; y^n ) + I(x^n \rightarrow y^n || Dx^n ) 
\label{eq:conservationLaw}
\end{eqnarray}
where 
\begin{eqnarray}
 I(x^n \rightarrow y^n || Dx^n )  &=&  \sum_i I(x^{i};y_{i}|y^{i-1},x^{i-1}) \nonumber \\
 & =& \sum_i I(x_{i};y_{i}|y^{i-1},x^{i-1}) 
\end{eqnarray}
 This proves 
$I(x^n \rightarrow y^n ) + I(y^n \rightarrow x^n ) $ is
symmetrical but is in general not equal to the mutual information, 
except if and only if  $ I(x_{i};y_{i}|y^{i-1},x^{i-1})=0, \forall i=1,\dots,n$.  Since the term in the sum is the mutual information between the present samples of the two processes conditioned on their joint past values, this measure is a measure of instantaneous dependence.
The term 
$I(x^n \rightarrow y^n || Dx^n )  = I(y^n \rightarrow x^n || Dy^n ) $ will thus be named the {\em instantaneous information exchange}  between   $x$ and $y$. 
%%%%%%%%%%%%%%% 

\subsection{Directed information rates}

Entropy as well as mutual information are  extensive quantities, increasing (in general) linearly with 
the length $n$ of the recorded time series.  
 Shannon's information rate for stochastic processes compensates the linear growth by considering  $A_\infty(x)= \lim_{n\rightarrow+\infty} A_n(x) / n $ ( if the limit exists), where $A_n(x)$
denotes any information measure on the sample $x$ of length $n$. 
 
For the important class of stationary processes (see e.g. \cite{CoveT93})  the entropy rate turns out to be the limit of the conditional entropy~: 
\begin{eqnarray}
\lim_{n\rightarrow+\infty}\frac{1}{n} H(x^n)  =  \lim_{n\rightarrow+\infty} H(x_{n}| x^{n-1})
\end{eqnarray}
 Kramer generalized this result for causal conditional entropies, thus defining 
 the directed information rate for stationary processes  as
 \begin{eqnarray}
I_\infty(x\rightarrow y) &=& \lim_{n\rightarrow+\infty} \frac{1}{n}\sum_{i=1}^{n} I( x^{i} ; y_i |y^{i-1}) \nonumber \\
&=& \lim_{n\rightarrow+\infty} I( x^n ; y_n |y^{n-1})
\label{eq:rateDirInfo}
\end{eqnarray}
This result holds also for the instantaneous information exchange rate. 
 Note that the proof of the result relies on the positivity of the entropy    for discrete valued stochastic processes. 
For continously valued processes, for which entropy can be negative, the proof is more involved and requires the methods developed  in \cite{Pins64,GrayK80,Gray90}, see also \cite{TatiM09}. 

%%%%%%%%

\subsection{Transfer entropy and instantaneous information exchange}

Introduced by Schreiber in  \cite{Schr00,KaisS02}, {\em transfer entropy} evaluates the deviation of the observed data from a model assuming the following  joint Markov property
\begin{eqnarray}
p(y_n | y_{n-k+1}^{n-1}, x_{n-l+1}^{n-1} ) = p(y_n | y_{n-k+1}^{n-1})
\label{eq:jointMarkov}
\end{eqnarray}
This leads to the following definition
\begin{eqnarray}
T(x_{n-l+1}^{n-1}\rightarrow y_{n-k+1}^{n})  = E\left[  \log \frac{p(y_n | y_{n-k+1}^{n-1}, x_{n-l+1}^{n-1} ) }{p(y_n | y_{n-k+1}^{n-1})} \right]
\end{eqnarray}
Then $T(x_{n-l+1}^{n-1}\rightarrow y_{n-k+1}^{n}) =0 $ iff eq. (\ref{eq:jointMarkov}) is satisfied.
Although in the original definition the past of $x$ in the conditioning may  begin at a different time $m \not=n$,  for practical reasons $m=n$ is considered. Actually, no {\it a priori } is available about possible delays, and setting $m=n$ allows to compare the  transfer entropy with the directed information. 

 By expressing  the transfer entropy as a difference of conditional entropies, we get 
\begin{eqnarray}
T(x_{n-l+1}^{n-1} \rightarrow  y_{n-k+1}^{n} )  &= & H(y_n | y_{n-k+1}^{n-1}) - H(y_n |y_{n-k+1}^{n-1}, x_{n-l+1}^{n-1} ) \nonumber\\
&=& I(x_{n-l+1}^{n-1} ; y_n | y_{n-k+1}^{n-1})
\end{eqnarray}
For $l=n=k$,  the identity
 $I(x,y ; z|w)=I(x;z|w)+I(y;z|x,w)$  leads to
 \begin{eqnarray}
 I( x^n ; y_n |y^{n-1}) &= I( x^{n-1} ; y_n |y^{n-1}) + I(x_n ; y_n |x^{n-1} , y^{n-1}) \nonumber \\
	 &= T(x^{n-1} \rightarrow  y^n ) +  I(x_n ; y_n |x^{n-1} , y^{n-1})
	 \label{eq:SchreiberT}
\end{eqnarray}
For stationary processes, letting $n\rightarrow \infty$ and  provided the limits exist, we obtain for the rates
 \begin{eqnarray}
I_\infty( x \rightarrow y) = T_\infty(x \rightarrow y ) +  I_\infty(x \rightarrow y  || 	Dx )
\end{eqnarray}
Transfer entropy is the part of the directed information that measures the causal influence of the past of $x$ onto the present of $y$. However it does not take into account the possible instantaneous  dependence of one time series on another, which is handled by directed information. 

Moreover,  only $ I( x^{i-1} ; y_i |y^{i-1})$ is considered in $T$, instead of  its sum over $i$  in the directed information. Thus stationarity is implicitly assumed and the transfer entropy has  the same 
meaning as  a rate. Summing 
over $n$ in eq. (\ref{eq:SchreiberT}),  the following  decomposition of the directed information is obtained 
\begin{eqnarray}
I(x^n \rightarrow  y^n ) = I( Dx^n \rightarrow  y^n ) + I(x^n \rightarrow  y^n  || Dx^{n} )
\label{dirinfodecomp:eq}
\end{eqnarray}
Eq. (\ref{dirinfodecomp:eq}) establishes that the influence of one process on another may be decomposed into two terms accounting for the past and for instantaneous contributions respectively.

%%%%%
%toto
\subsection{Accounting for side information}
The preceding definitions all aim at proposing definitions of information exchange between $x$ and $y$;  the possible information gained from possible connections with the rest of the network is not taken into account.  The other possibly observed time series are hereafter referred to as side information. The available side information at time $n$ is noted $z^n$. Then, two conditional quantities are introduced~: conditional directed information and causally conditioned directed information. 
\begin{eqnarray}
I(x^n \rightarrow  y^n \big| z^{n} ) &= & H\big(y^n\big| z^{n} ) - H\big(y^n \big|\big|  x^n\big| z^{n}\big) \\
I(x^n \rightarrow  y^n \big|  \big|z^{n} ) &= & H\big(y^n\big| \big| z^{n} ) - H\big(y^n \big|\big|  x^n, z^{n}\big) 
\end{eqnarray}
 where
\begin{eqnarray}
H\big( y^n \big| x^n \big|\big| z^{n} \big)&=  &H\big( y^n, x^n \big|\big| z^{n} \big)-H\big(  x^n \big|\big| z^{n} \big) \\
H\big(y^n \big|\big|  x^n \big| z^{n} \big)   &= &  \sum_{i=1}^n H\big(y_{i} \big| y^{i-1} , x^{i}, z^{n} \big)
\end{eqnarray}
 In these equations, following \cite{Kram98}, conditioning goes from left to right~:  the first conditioning type met is the one applied. \\
Note that for usual conditioning, variables do not need to be synchronous with the others and can have any dimension. The synchronicity  constraint appear in the new definitions above.

%%%%%%%%%%%%%%%%%%%%%
For conditional directed information, 
a conservation law similar to eq. (\ref{eq:conservationLaw}) holds:
\begin{eqnarray}
I(x^n \rightarrow y^n \big| z^{n} ) +  I( Dy^n \rightarrow x^n  \big| z^{n} )&=& I( x^n ; y^n  \big| z^{n} )  
\end{eqnarray}
Furthermore,  
conditional mutual and directed information are equal if and only if 
\begin{eqnarray}
H(x_i \big| x^{i-1},y^{i-1}, z^{n}) = H(x_i \big| x^{i-1}, z^{n} ) \mbox{ } \forall i=1,\ldots,n
\end{eqnarray}
 which means that given the whole observation of the side information, there is no feedback from $y$ to $x$.  
 Otherwise stated, if there is feedback from $y$ to $x$ and if $I( Dy^n \rightarrow x^n  \big| z^{n} )=0$, the feedback from $y$ to $x$ goes  through $z$. 

Finally, let us  mention that conditioning with respect to some stationary time series $z$ similarly leads to define the causal directed information rate as
\begin{eqnarray}
I_\infty(x\rightarrow y \big| \big| z) &=& \lim_{n\rightarrow+\infty} \frac{1}{n}\sum_{i=1}^{n} I( x^{i} ; y_i |y^{i-1}, z^{i}) \\
&=& \lim_{n\rightarrow+\infty} I( x^n ; y_n |y^{n-1} , z^{n})
\end{eqnarray}

This concludes the presentation of directed information. We have put emphasis on the importance of Kramer's causal conditioning, both for the definition of directed information and  for taking into account  side information.  We have also proven that Schreiber's transfer entropy is that part of directed information dedicated to the strict sense causal information flow (not accounting for simultaneous coupling). Next section revisits Granger causality as another means for assessing influences between time series.

\section{Granger causality between multiple time series}
\label{granger:sec}

\subsection {Granger's definition of causality}

Although no universally well accepted definition of causality exists, Granger approach of causality  is often preferred  for two major reasons. The first reason is to be found in the apparent simplicity of the definitions and axioms proposed in the early papers, that suggests  the following probabilistic approach for causality~: $y_n$ is said to cause $x_{n}$ if 
\begin{eqnarray}
 \mbox{ Prob } (x_{n}\in A|\Omega_{n-1})\neq \mbox{Prob } (x_{n}|\Omega_{n-1}\backslash y^{n-1})
\end{eqnarray} 
for any subset $A$.  $\Omega_{n}$ was called by Granger  {\em  ``all the information available in the universe''} at time $n$, whereas $\Omega_n\backslash y^n$ stands for all information except $y^n$. In practice, $\Omega_n\backslash (x^n,y^n)$ is the side information $z^n$.

The second reason is that, in his 1980 paper \cite{Gran80}, Granger introduced a set of operational definitions, thus allowing to derive practical testing procedures. These procedures require the introduction of models for testing causality, although Granger's approach and definitions are fully general; furthermore, Granger's approach raises the important issues below: 
\begin{enumerate}
\item Full causality is expressed in terms of probability and leads to relationships between probability density functions.  Restricting causality to relations defined on mean quantities is less stringent and allows more practical approaches.
\item Instantaneous dependence may be added to  the causal relationships, {\it e.g. } by adding $y_{n}$ to the set of observations available at time $n-1$. This leads to a weak concept as it is no longer possible to discriminate between instantaneous causation of $x$ by $y$, of $y$ by $x$ or of feedback, at least without  imposing extra structures to the data models. 
\item It assumed that $\Omega_{n}$ is separable~:  $\Omega_n \backslash y^n$ must be defined. This point is crucial  for practical issues: the causal relationship between $x^n$ and $y^n$ (if any)  is intrinsically related to the set of available knowledge at time $n$. Adding a new subset of observations, {\it e.g.} a new time series, may lead to different conclusions  when testing causal dependencies between $x$ and $y$.
\item If $y_n$ is found to cause $x_{n}$ with respect to some observation set, this does not preclude the possibility that $x_n$ causes $y_{n}$ if there exists some feedback between the two series. 
\end{enumerate}
Item 2 above motivated Geweke's approaches \cite{Gewe82,Gewe84}, discussed below. Item 3 and 4 highlights the importance of conditioning the information measures to the set of available observations (related to nodes that may be connected to either $x$ or $y$), in order to identify causal information flows between any pair of nodes in a multi-connected network. As a central purpose of this paper is to relate Granger causality and directed information in presence of side information, the practical point of view suggested by Geweke is adopted. It consists in introducing a linear model for the observations.

%%%%%%%%%%%%%%%%%%%%%%%%%%%%%
%%%%%%%%%%%%%%%%%%%%%%%%%%%%%
%%%%%%%%%%%%%%%%%%%%%%%%%%%%%

\subsection{Geweke's approach}
Geweke proposed measures of (causal) linear dependencies and feedback  between two multivariate Gaussian time series $x$ and $y$. 
A third time series $z$ is introduced as side information. This series allows to account for the influence of other nodes interacting with either $x$ or $y$, as this may  happen in networks where many different time series or multivariate random processes are recorded.  
The following parametric model is assumed, 
\begin{eqnarray}
 \left\{\begin{array}{lcl}
   x_n  &=&\displaystyle   \sum_{s=1}^\infty A_{i,s} x_{n-s}  + \sum_{s=0}^\infty B_{i,s} y_{n-s} +\sum_{s=0}^b\alpha_{i,s}z_{s} + u_{i,t} \\
     y_n  & =&\displaystyle \sum_{s=0}^\infty C_{i ,s} x_{n-s}  + \sum_{s=1}^\infty D_{i,s} y_{n-s} + \sum_{s=0}^b\beta_{i,s}z_{s}+ v_{i,t} 
\end{array}\right.
\label{modeles:eqbis}
\end{eqnarray}
Accounting for   all $z$  corresponds to $b=\infty$ in eq. (\ref{modeles:eqbis}), whereas causally conditioning on $z$ is obtained by setting $b=n-1$.
Furthermore, it is assumed that all  the  processes studied are jointly Gaussian. Thus the analysis can be restricted to second order statistics only. 

Under the assumption that  the coefficients $\alpha_{i,s}$ and $\beta_{i,s}$ are set to zero (leading back to original Geweke's model), we easily see that  eq. (\ref{modeles:eqbis}) can actually handle three different dependence models indexed by $i=\{1,2,3\}$, as defined below~: 
\begin{itemize}
\item $i=1$~:  no coupling exists,   $B_{1,s}=0, C_{1,s}=0,  \forall s$  and the prediction residuals $u_{1,t}$ and $v_{1,t} $ are white and independent random processes. 
\item $i=2$, both series are only dynamically coupled~:  $B_{1,0}=0, C_{1,0}=0$ and the prediction residues are white random processes. Linear prediction properties allow to show  that the cross correlation function of  $u_{2,t}$ and $v_{2,t}$ is different from zero for the null delay only~: $\Gamma_{2,uv}(t) = \sigma^2 \delta(t)$.
\item $i=3$;  the time series are  coupled ~: $B_{3,s}\not=0, C_{3,s}\not=0, \forall s$ and the residues $u_{3,t}$ and $v_{3,t} $ are white,  but are no longer independent. 
\end{itemize} 
Note that models 2 and 3 differ only by the presence (model 3) or absence (model 2) of instantaneous coupling. It can be shown that these models are `equivalent' if $\sigma^{2} \neq 0$, thus allowing to compute an invertible linear mapping  that transforms model 2 into a model of type 3. This confirms that model 3 leads to some weak concept, as already quoted previously. The same analysis and conclusions hold when the coefficients  $\alpha_{i,s}$ and $\beta_{i,s}$ are restored. 

\subsection{Measures of dependence and feedback. } 
 Geweke \cite{Gewe82,Gewe84}  introduced dependence measures constructed on the covariances of the residues $u_{i,t}$ and $v_{i,t} $ in (\ref{modeles:eqbis}). We briefly recall these measures. Let 
\begin{eqnarray}
\varepsilon_\infty^2(x_n| x^{n-1}, y^{l},z^{b})= \lim_{n\rightarrow +\infty} \varepsilon^2(x_n| x^{n-1}, y^{l},z^{b})
\end{eqnarray}
for   $l=n$ or $n-1$ according to the considered model.

$\varepsilon_\infty^2(x_n| x^{n-1}, y^{l},z^{b})$ is the asymptotic variance\footnote{The presence of $n$ in the notation $\varepsilon_\infty^2()$ is  an abuse of notation, but is adopted to keep track of the variables involved in this one-step forward prediction.} 
of the prediction residue when predicting $x_n$ from the observation of  $x^{n-1}$ , $ y^{l}$ and $ z^{b}$. For multivariate processes, $\varepsilon^2()$ is given by the determinant $\det \Gamma_{i,()}$ of the covariance matrix of the residues. 

 Depending on the value of $b$ in (\ref{modeles:eqbis}), and following Geweke, the following measures are proposed for $b=\infty$~: 
\begin{eqnarray}
F_{y\rightarrow x|z}& =& \log \frac{\varepsilon_\infty(x_n| x^{n-1},z^{\infty})}{\varepsilon_\infty(x_n| x^{n-1}, y^{n-1},z^{\infty})} \nonumber \\
F_{x\rightarrow y|z}& = &\log  \frac{\varepsilon_\infty(y_n| y^{n-1},z^{\infty})}{\varepsilon_\infty(y_n| x^{n-1}, y^{n-1},z^{\infty})} \nonumber \\
 F_{x.y|z} &= &\log  \frac{\varepsilon_\infty(x_n| x^{n-1},y^{n-1},z^{\infty})}{\varepsilon_\infty(x_n| x^{n-1}, y^n,z^{\infty})} 
\end{eqnarray}
and for causal conditioning, $b=n-1$: 
\begin{eqnarray}
F_{y\rightarrow x||z}& =& \log \frac{\varepsilon_\infty(x_n| x^{n-1},z^{n-1})}{\varepsilon_\infty(x_n| x^{n-1}, y^{n-1},z^{n-1})}\nonumber \\
F_{x\rightarrow y||z}& = &\log  \frac{\varepsilon_\infty(y_n| y^{n-1},z^{n-1})}{\varepsilon_\infty(y_n| x^{n-1}, y^{n-1},z^{n-1})} \nonumber \\
 F_{x.y||z} &= &\log  \frac{\varepsilon_\infty(x_n| x^{n-1},y^{n-1},z^{n-1})}{\varepsilon_\infty(x_n| x^{n-1}, y^n,z^{n-1})} 
\end{eqnarray}
Note that these measures are greater or equal to zero. 

Remarks~: 
\begin{itemize}
  \item $F_{x.y|z}  $ and $F_{x.y||z}  $ can be shown to  symmetric with respect to $x$ and $y$  \cite{Gewe82,Gewe84}.  This is not the case for the other measures: if strictly positive, they  indicate  a direction in the coupling relation.
  
  \item  Causally conditional on $z$, $F_{x\rightarrow y||z} $ measures the linear feedback   from $x$ to $y$ and $F_{x.y||z}  $ measures the  instantaneous linear feedback, as  introduced by Geweke.

  \item In \cite{RissW87},   Rissanen and Wax introduce measures which are no longer constructed from the variances of the prediction residues but rather from a quantity of information (measured in bits) that is required for performing linear prediction. One cannot afford to deal with infinite order in the regression models, and these approaches are equivalent to Geweke's. In \cite{RissW87}, the information contained in the model order selection is taken into account. We will not develop this aspect in this paper. 
\end{itemize}

\section{Directed information and Granger Causality}
\label{DirInfoGranger:sec}
We begin by studying the linear Gaussian case, and close the section by a more general discussion.

\subsection{Gaussian linear models}
Although it is not fully general,  the Gaussian case allows to develop
  interesting insights into directed information. Furthermore,  it provides a bridge between directed information theory and causal inference in networks, as partly described in an earlier work \cite{AmblM09}, \cite{BarnBS09}. The calculations below are conducted without taking observations others than $x$ and $y$, as it is straightforward to generalize in the presence of side information.

Let $H(y^{k}) = 1/2\log (2\pi e)^k |\det \, \Gamma_{y^{k}} |$ be the entropy of the $k$ dimensional  Gaussian random vector $y^{k}$ of covariance matrix $\Gamma_{y^{k}}$.
Using block matrices properties, we have
\beq
\det \, \Gamma_{y^{k}} = \varepsilon^2(y_k|y^{k-1}) \det \, \Gamma_{y^{k-1}} 
\label{eq:RecursivGamma}
\eeq
where $\varepsilon^2(y_k|y^{k-1})$ is the linear prediction error of $y$ at time $k$ given its past \cite{BrocD91} .
Then, the entropy increase is\footnote{Note that if $y$ is a stationary stochastic process, the limit of the 
 entropy difference in eq. (\ref{eq:entropyDiff})
 is nothing but the entropy rate. Thus, taking the limit of eq. (\ref{eq:entropyDiff}) exhibits the well known
relation between  entropy rate and asymptotic one step linear prediction \cite{CoveT93}.}
\begin{eqnarray}
H(y^{k}) -H(y^{k-1}) &=&\frac{1}{2} \log\Big|\frac{\det \,\Gamma_{y^{k}}}{\det \,\Gamma_{y^{k-1}}}\Big|  \\
&=& \frac{1}{2} \log  \varepsilon^2(y_k |  y^{k-1})
\label{eq:entropyDiff}
\end{eqnarray}
%%%%%%%%%%%%%%%%%%%%%%%%%%%%%%%%%%%
Let $ \varepsilon^2(y_k| y^{k-1}, x^{k})$ be the power of the linear estimation error of $y_k$ given its past and  the observation of $x$ up to time $k$. 
Since  
$
I( x^{k} ; y_k  \big| y^{k-1})
= H(y^{k}) -H(y^{k-1})  - H( x^{k} ;y^{k}) +  H( x^{k} ;y^{k-1}) 
$,  the conditional mutual information and the directed mutual information respectively writes
\begin{eqnarray}
\label{eq:condMutInf}
I( x^{k} ; y_k  \big| y^{k-1}) &=& \frac{1}{2} \log \frac{\varepsilon^2(y_k |  y^{k-1})}{ \varepsilon^2(y_k| y^{k-1}, x^{k})} \\
I(x^n\rightarrow y^n) &=& \frac{1}{2}\sum_{i=1}^n \log \frac{\varepsilon^2(y_i |  y^{i-1})}{ \varepsilon^2(y_i| y^{i-1}, x^{i})}
\end{eqnarray}

If furthermore the vectors considered above are built from  
jointly stationary Gaussian processes, 
letting $n\rightarrow \infty$ in eq. (\ref{eq:condMutInf}) gives   the directed information rates:
\begin{eqnarray}
 I_\infty(x\rightarrow y) &=&  \frac{1}{2} \log \frac{\varepsilon^2_\infty(y_k |  y^{k-1})}{ \varepsilon^2_\infty(y_k| y^{k-1}, x^{k})}
 \label{eq:IInfinite}
\end{eqnarray}
where $\varepsilon^2_\infty(y_k |  y^{k-1})$ is the asymptotic power of the one step linear prediction error.
By reformulating eq. (\ref{eq:IInfinite}) as
\begin{eqnarray}
\varepsilon^2_\infty(y_k| y^{k-1}, x^{k}) = e^{-2  I_\infty(x\rightarrow y)  } \varepsilon^2_\infty(y_k |  y^{k-1})
\end{eqnarray}
shows  that the directed information rate measures the advantage of including the process $x$ into the prediction of process $y$. 

If side information  is available as a time series $z$, and  
if  $x$, $y$ and $z$ are jointly stationary,  
the same arguments as above lead to 
\begin{eqnarray}
\varepsilon^2_\infty(y_k| y^{k-1}, x^{k}, z^{k-1}) = e^{-2  I_\infty(x\rightarrow y || Dz)  }
 \varepsilon^2_\infty(y_k |  y^{k-1}, z^{k-1})
\end{eqnarray}
where recall that $Dz$ stands for the delayed time series.
This equation  
highlights that causal conditional directed information has the same meaning as directed information, provided that we are measuring the information gained by considering $x$ in the prediction of $y$ given its past and the past of $z$.

\subsection{ Relations between Granger and Massey's approaches}

To relate previous results to  Granger causality, the contribution of the past values must be separated from those related to instantaneous coupling in the directed information expressions. 
A natural framework is provided by transfer entropy. 

From eq. (\ref{dirinfodecomp:eq}),  under the assumption that the studied processes are jointly Gaussian, arguments similar to those used in the previous paragraph lead to
\begin{eqnarray}
I(Dx^n\rightarrow y^n) = \frac{1}{2}\sum_{i=1}^n \log \frac{\varepsilon^2(y_i |  y^{i-1})}{ \varepsilon^2(y_i| y^{i-1}, x^{i-1})}  \\
I(x^n\rightarrow y^n || Dx^n)  =  \frac{1}{2}\sum_{i=1}^n \log \frac{\varepsilon^2(y_i |  y^{i-1}, x^{i-1})}{ \varepsilon^2(y_i|y^{i-1}, x^{i})}  
\end{eqnarray}
Likewise, causally conditioned directed information decomposes as 
\begin{eqnarray}
I(x^n\rightarrow y^n|| Dz^{n})= I( Dx^n \rightarrow y^n || Dz^{n} ) + I(x^n \rightarrow y^n || Dx^n , Dz^{n}) 
\end{eqnarray}
 Expressing conditional information as a function of prediction error variance we get
\begin{eqnarray*}
I(x^n\rightarrow y^n|| Dz^{n}) =  \frac{1}{2}\sum_{i=1}^n \log \frac{\varepsilon^2(y_i |  y^{i-1}, z^{i-1})}{ \varepsilon^2(y_i| y^{i-1}, x^{i-1}, z^{i-1})} + 
 \frac{1}{2}\sum_{i=1}^n  \log \frac{\varepsilon^2(y_i |  y^{i-1}, x^{i-1},z^{i-1})}{ \varepsilon^2(y_i|y^{i-1}, x^{i} ,z^{i-1})}
\end{eqnarray*}
The first term accounts for the influence of the past of $x$ onto $y$, whereas the second term evaluates the instantaneous influence of $x$ on $y$, provided $z$ is (causally) observed.

Finally, letting $n \rightarrow \infty$, the following relations between directed information measures and generalized Geweke's indices  are obtained~: 
\begin{eqnarray*}
I_\infty (Dx \rightarrow y ) &= &\frac{1}{2} \log \frac{\varepsilon_\infty^2(y_i |  y^{i-1})} { \varepsilon^2(y_i| y^{i-1}, x^{i-1})}  = F_{x\rightarrow y} \\
I_\infty(x \rightarrow y  || Dx ) & =&  \frac{1}{2}  \log \frac{\varepsilon_\infty^2(y_i |  y^{i-1}, x^{i-1})}{ \varepsilon^2(y_i|y^{i-1}, x^{i})}   = F_{x.y}\\
I_\infty( Dx  \rightarrow y  || z  )&=& \frac{1}{2}  \log \frac{\varepsilon_\infty^2(y_i |  y^{i-1}, z^{i-1})}{ \varepsilon^2(y_i| y^{i-1}, x^{i-1}, z^{i-1})} =F_{x \rightarrow y||z} \\ 
 I_\infty(x  \rightarrow y  || Dx  , z )&=&\frac{1}{2}   \log \frac{\varepsilon_\infty^2(y_i |  y^{i-1}, x^{i-1},z^{i-1})}{ \varepsilon^2(y_i|y^{i-1}, x^{i} ,z^{i-1})}=F_{x. y||z} 
\end{eqnarray*}
 This proves that for Gaussian processes, directed information rates (causal conditional or not) and Geweke's indices are in perfect match.  

\subsection{Directed Information as a generalized Granger's approach}
\label{defproposal:ssec}

These results are obtained under Gaussian assumptions and are closely related to linear prediction theory. However,  the equivalence between Granger's approach and directed information can hold  in a more general framework by proposing the following information theoretic based definitions of causal dependence: 
\begin{enumerate}
\item $x_t$ is not a cause of $y_t$  with respect to $z_t$ if and only if $I_\infty( Dx  \rightarrow y  || Dz  ) = 0 $
\item $x_t$ is not instantaneously causal to $y_t$  with respect to $z_t$ if and only if $ I_\infty(x  \rightarrow y  || Dx  , Dz )=0$
\end{enumerate}
These directed information based definitions generalize Granger's approach. Furthermore, these new definitions allow to infer graphical models for multivariate time series~: This builds a strong connection between the present framework and Granger causality graphs developed by Eichler and Dalhaus \cite{DahlE03}. 
This connection is further explored in \cite{AmblM11} and in the recent work \cite{QuinKC11}.

\section{Application to multivariate Gaussian processes}
\label{threeprocesses:sec}

To illustrate the preceding results, we study the information flow between components of a multivariate Gaussian process. To stress the importance of causal conditioning and of availability of side information, we separate the bivariate analysis from the multivariate analysis. Furthermore, we particularly concentrate on a first order  autoregressive model.

Let $X_n=C X_{n-1} +W_n$ be a multidimensional stationary, zero-mean, Gaussian process. $W_n$ is  a Gaussian white multidimensional noise with correlation matrix $\Gamma_w$ (not necessarily  diagonal).  The off-diagonal terms in matrix $C$ describe the interactions between the  components of $X$.
$c_{ij}$  denotes the coupling coefficient from component $i$ to component $j$.
 The correlation matrix of $X$ is a solution of the equation  
\beq
\Gamma_{X} = C \Gamma_X C^t +\Gamma_w
\label{eq:Gamma}
\eeq
Main directed information measures are firstly evaluated on a bivariate process. Then side information is  assumed to be observed, and the same information measures are reconsidered for different coupling models.

\subsection{Bivariate AR(1) model}

Let  $[v_n, w_n]^t = W_n$ and  $\sigma_v$ , $\sigma_w$ be their standard deviations and $\gamma_{vw}$ their correlation coefficient.  
Let $X_n = [x_n,y_n]^t$.  
$\Gamma_{X}$ is computed by solving eq. (\ref{eq:Gamma}) as a function of the coupling  coefficients between $x_n$ and $y_n$. The initial condition $(x_1,y_1)$ is assumed to follow the same distribution as  $(x_n,y_n)$ to ensure the absence of transients. 

Under these assumptions, some computations lead to express  the mutual and directed information as
\begin{eqnarray}
I(x^n; y^n)&=& \frac{n-1}{2} \log\left(  \frac{( c_{yx}^2  \sigma_y^2 +\sigma_v^2)( c_{xy}^2  \sigma_x^2 +\sigma_w^2)}{ \sigma_v^2 \sigma_w^2+\gamma_{vw}^2} \right) + I(x_1;y_1) \label{eq:MI}\\
I(x^n \rightarrow  y^n) &=& \frac{n-1}{2} \log \left(   \frac{c_{xy}^2  \sigma_x^2 +\sigma_w^2}{\sigma_v^2 \sigma_w^2-\gamma_{vw}^2} \right) + I(x_1;y_1) + \frac{n-1}{2}\log\left(    \sigma_v^2 \right) \label{MId} \\
I(y^n \rightarrow  x^n) &=& \frac{n-1}{2} \log \left(  \frac{c_{yx}^2  \sigma_y^2+\sigma_v^2}{\sigma_v^2 \sigma_w^2-\gamma_{vw}^2} \right) + I(x_1;y_1) + \frac{n-1}{2}\log\left( \sigma_w^2 \right) \label{MIr}
\end{eqnarray}
where $I(x_1;y_1)=-1/2 \log\big( 1-\gamma_{xy}^2/ (\sigma_x^2  \sigma_y^2 ) \big) $ and $\gamma_{xy}$ stands for the correlation between $x$ and $y$. 

Equations (\ref{eq:MI},\ref{MId},\ref{MIr}) raise some comments: 
 \begin{enumerate}
  \item The directed information is clearly asymmetric.
  \item On one hand, the left hand side of the conservation equation (\ref{eq:conservationLaw})  is given by summing equations (\ref {MId})  and (\ref {MIr}). On the other hand summing the mutual information (\ref{eq:MI}) and  
  \begin{eqnarray}
I(x^n \rightarrow y^n|| Dx^n ) &=& I(x_1;y_1) +\sum_{i \geq 2} I(x_{i},y_i | y^{i-1}, x^{i-1} ) \\
 &=& \frac{n-1}{2}\log \left(  \frac{ \sigma_v^2 \sigma_w^2}{\sigma_v^2 \sigma_w^2-\gamma_{vw}^2} \right) +I(x_1;y_1)
\end{eqnarray} 
gives as expected  the right hand side of the conservation equation (\ref{eq:conservationLaw}). We recover the fact that for independent noise components ($\gamma_{vw}=0$) 
the sum of the directed information flowing in opposite directions is equal to the mutual information. 
This is however not the case in general. 
  \item The information rates are obtained by letting $n\rightarrow \infty$ in eq. (\ref{MId}), (\ref{MIr}):
\begin{eqnarray}
I_\infty(x  \rightarrow  y ) &=& \frac{ 1}{2} \log \left( \frac{c_{xy}^2  \sigma_x^2 +\sigma_w^2}{\sigma_v^2 \sigma_w^2-\gamma_{vw}^2} \times \sigma_v^2  \right) \\
I_\infty(y  \rightarrow  x ) &=& \frac{ 1}{2} \log \left( \frac{c_{yx}^2  \sigma_y^2+\sigma_v^2}{\sigma_v^2 \sigma_w^2-\gamma_{vw}^2} \times \sigma_w^2 \right)  
\end{eqnarray}
This shows that if {\it e.g. } $c_{yx}=0$, 
we observe that  a coupling is equal to zero in one direction, the directed information rate from $y$ to $x$ 
satisfies   
 \begin{eqnarray}
\frac{ 1}{2} \log \left( \frac{ \sigma_w^2 \sigma_v^2}{\sigma_v^2 \sigma_w^2-\gamma_{vw}^2}    \right) 
 = \lim_{n\rightarrow +\infty} \frac{1}{n} I(x^n \rightarrow y^n|| Dx^n)
 \end{eqnarray}
The right hand side of the equality  
may be interpreted as a lower bound for the directed informations rates. In particular, when 
$\Gamma_w$ is diagonal, this bound is zero.

This corresponds to the decomposition (\ref{dirinfodecomp:eq}) for the rates, 
$I_\infty(x  \rightarrow  y ) = I_\infty(Dx  \rightarrow  y ) + I_\infty(x  \rightarrow  y || Dx) $. The first term 
$ I_\infty(Dx  \rightarrow  y ) = \frac{ 1}{2} \log \left( 1+ \frac{c_{xy}^2  \sigma_x^2}{\sigma_w^2} \right)$ 
 is 
 Schreiber's transfer entropy, or the directed information from the past of $x$ to $y$ (this term is equal to the first index of Geweke in this case). 
The second term $I_\infty(x  \rightarrow  y || Dx) $ 
 corresponds to the second of Geweke's indices and 
measures the instantaneous coupling between the time series.  
%%%%%%%%%%%%%%%%%%%%
 \item Directed information increases with the coupling strength, as expected for a measure of information flow. 
  \end{enumerate}

\subsection{Multivariate AR(1) model}
\begin{figure}[p]
    \begin{center}
      \includegraphics{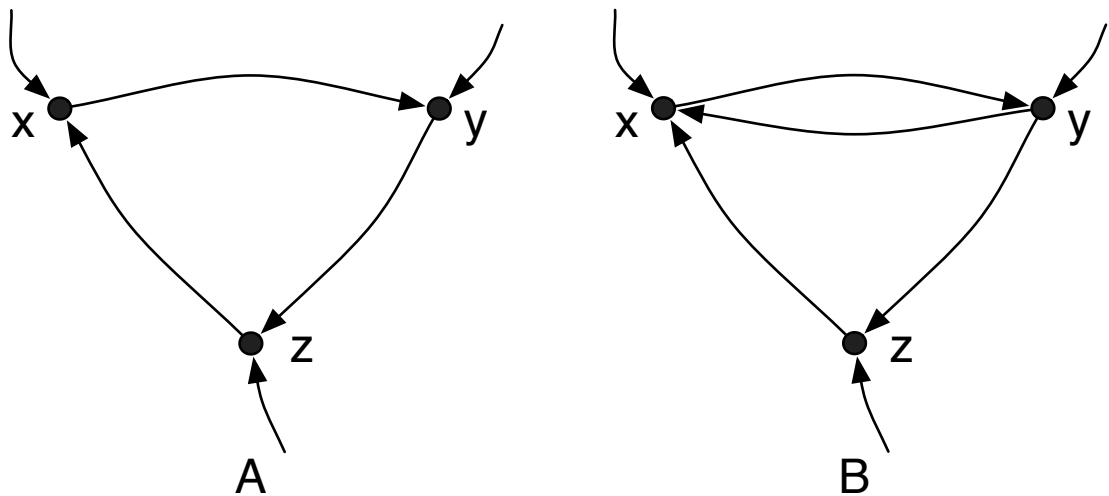}
    \end{center}
    \caption{ Networks of three Gaussian processes studied in the paper. An arrow represents a  coupling coefficient not equal to zero from the past of one signal to the other. In frame A, there is no direct feedback between any of the signals. However, a feedback from $y$ to $x$ exists through $z$. In frame B, there is also a direct feedback from $y$ to $x$. The arrows coming from the outside of the network represent the inputs, {\it i.e. } the dynamical noise $W$ in the AR model. }
    \label{graph:fig}
    \end{figure}

 Let $X_n=[z_n, x_n, y_n]^t$, $W_n = [u_n, v_n, w_n]^t$ be  a three dimensional Gaussian stationary zero mean process satisfying the AR(1) equation, satisfying the same set of hypothesis and notations  as above. We study two cases described in figure (\ref{graph:fig}), 
 where the arrows indicate the coupling direction. 
 
The distributions of the variables 
$y_n|y^{n-1}$ and $y_n| y^{n-1},x^{n-1}$ required in the calculation of {\it e.g. } $I_\infty(x \rightarrow y)$ are difficult to obtain explicitly. Actually, even if $X$ is a Markov process, the components are not. However since we deal with and AR(1) process,  $p(y_n | X^{n-1}) =p(y_n| X_{n-1})$ and  $p(x_n, y_n | X^{n-1}) =p(x_n,y_n| X_{n-1})$. The goal is to evaluate $I_\infty(y\rightarrow x || Dz) $.
As $I(y^n ; x_n | x^{n-1}, z^{n-1}) = I(y^{n-1} ; x_n | x^{n-1}, z^{n-1}) + I(y_n ; x_n | X^{n-1})$, one has 
\begin{eqnarray}
I_\infty(y\rightarrow x || Dz) &= & \lim_{n\rightarrow \infty} I(y^n ; x_n | x^{n-1}, z^{n-1})\nonumber \\ 
 &=&\lim_{n\rightarrow \infty} I(y^{n-1} ; x_n | x^{n-1}, z^{n-1}) -1/2\log(1-\gamma_{vw}^2/(\sigma_v^2 \sigma_w^2))
\end{eqnarray}
where $\gamma_{vw}$ is the correlation coefficient between $v_n$ and $w_n$.

In case B 
 (see figure \ref{graph:fig}), there is feedback from $y$ to $x$. 
Since  conditioning is over the past of $x$ and $z$ and since there is no feedback from $z$ to $y$, $x_n|(x,z)^{n-1}$ is normally distributed with variance $c_{yx}^2 \sigma_y^2 +\sigma_v^2$. Thus, we obtain 
for this case
\begin{eqnarray*}
I_{B,\infty}(y\rightarrow x || Dz) = \frac{1}{2}\log(1+ \frac{c_{yx}^2 \sigma_y^2}{\sigma_v^2} )- \frac{1}{2}\log(1+ \frac{\gamma_{vw}^2}{\sigma_v^2 \sigma_w^2} ) 
\end{eqnarray*}
Setting $c_{yx}=0$ we get  for case A, 
\beq 
I_{A,\infty}(y\rightarrow x || Dz) = -(1/2) \log(1-\gamma_{vw}^2/(\sigma_v^2 \sigma_w^2))
\eeq
which is  the instantaneous exchange rate between $x$ and $y$. 
 If  the noise components  $v$ and $w$ are independent, the causal conditional directed information is zero. 

\begin{figure}[p]
    \begin{center}
      \includegraphics{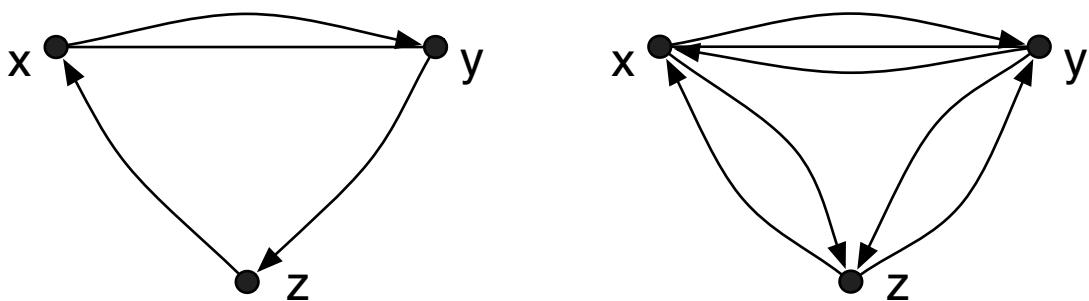}
    \end{center}
    \caption{ Networks of three Gaussian processes studied in the paper. The left plot corresponds to the correct model and to the inferred network when causal conditional directed information is used. The network on  the right is obtained if the analysis is only pairwise, when directed information is used between two signals without causal conditioning over the remaining signals. }
    \label{graph_infer:fig}
    \end{figure}

The preceding illustration 
 highlights the ability of causal conditioning to deal with different feedback scenarios in multiply connected stochastic networks. Figure \ref{graph_infer:fig} illustrates the inference result and the difference obtained if the third time series is not taken into account.

\section{Conclusion}

In this paper, we have revisited the directed information theoretic concept introduced by Massey, Marko and Kramer. A special attention has been paid to the key role played by causal conditioning. This turns out be be a central issue for characterizing information flows in the case where side information may be available. We propose a unified framework to enable a comparative study of mutual information, conditional mutual information with directed information in the context of networks of stochastic processes.  Schreiber's transfer  entropy, a widely used concept in physics and neuroscience,  is also shown to be easily interpreted with directed information tools.

The second section describes and discusses Granger causality  and its practical issues. Geweke's work serves as a reference in our discussion, and allows to provide a means to establish that Granger causality and directed information  lead to equivalent measures in the Gaussian linear case. Based upon the previous analysis, a possible extension of Granger causality definition is proposed. The extended definitions rely upon information theoretic criterion rather than probabilities, and allow to recover Granger's formulation in the linear Gaussian case. This new extended  formulation of Granger causality is of some practical importance for estimation issues. Actually, some recent works presented some advances in this direction; in \cite{QuinCKH11}, directed information estimators are derived from  spike trains models; in \cite{ViceWLP11}, Kraskov and Leonenko entropy estimators are used for estimating entropy transfer. In \cite{AmblM11}, the authors recourse to directed information in a graphical modeling context; Their equivalence with generative graphs for analyzing complex systems is studied in   \cite{QuinKC11}. The main contribution of the present paper is to provide a unified view that allow to recast causality and directed information within a unique framework. 

	Estimation issues were not mentioned in this study, as it may deserve a full paper {\it per-se}, and are deferred to a future work.

\section{Acknowledgements} P.O.A. is supported by an ExploraPro fellowship from R\'egion Rh\^one-Alpes and by a Marie Curie International Outgoing Fellowship from the European Community. We gratefully acknowledge
Pr. S. Sangwine (Univ. Essex, UK) for his thorough reading of the paper.

\bibliographystyle{unsrt}


\end{document}